%% file: EXO-10-014_temp.tex
\pdfoutput=1
\documentclass[11pt,twoside,a4paper,cmspaper,final,collab]{cms-tdr}
\def\svnVersion{29025:29032M}\def\cmsCernNo{2010-074}\def\cmsCernDate{\today}\def\cmsMessage{Submitted to Physics Letters B}

\begin{document}\cmsNoteHeader{EXO-10-014}
\hyphenation{env-iron-men-tal}
\hyphenation{had-ron-i-za-tion}
\hyphenation{cal-or-i-me-ter}
\hyphenation{de-vices}
\RCS$Revision: 29032 $
\RCS$HeadURL: svn+ssh://alverson@svn.cern.ch/reps/tdr2/papers/EXO-10-014/trunk/EXO-10-014.tex $
\RCS$Id: EXO-10-014.tex 29032 2010-12-29 13:32:47Z alverson $
\input{ptdr-definitions}
\providecommand\Wprime{\ensuremath{\mathrm{W'}}}
\providecommand\MT{\ensuremath{M_\mathrm{T}}}
\providecommand\invpb{$\mathrm{~pb}^{-1}\,$}

\providecommand\SM{SM\xspace}
\cmsNoteHeader{EXO-10-014} 
\title{Search for a heavy gauge boson W$^\prime$ in the final state with
  an electron and large missing transverse energy in pp collisions at
  $\sqrt{s} = 7$~TeV}
\author{The CMS Collaboration}
\date{\today}
\abstract{ A search for a heavy gauge boson \ensuremath{\mathrm{W'}}
  has been conducted by the CMS experiment at the LHC
  in the decay channel with an electron and large transverse
  energy imbalance $E_{\mathrm{T}}^{\text{miss}}$, using
  proton-proton collision data corresponding to an integrated
  luminosity of $36~\mathrm{pb}^{-1}$.  No excess above standard model
  expectations is seen in the transverse mass distribution of the
  electron-$E_{\mathrm{T}}^{\text{miss}}\xspace$ system.  Assuming
  standard-model-like couplings and decay branching fractions, a
  $\mathrm{W'}$ boson with a mass less than $1.36~\mathrm{TeV}/c^2$ is
  excluded at 95\% confidence level.  }
\hypersetup{%
pdfauthor={CMS Collaboration},%
pdftitle={Search for a heavy gauge boson W' in the final state with
  an electron and large missing transverse energy in pp collisions at
  sqrt(s) = 7 TeV},%
pdfsubject={CMS},%
pdfkeywords={CMS, physics, particle physics, LHC}}

\maketitle 

This letter describes a search for a heavy analogue of the standard model
$\PW$ gauge boson, \Wprime, where the particle decays
leptonically to an electron and a neutrino ($\Wprime \to \Pe \cPgn$).
Heavy partners of gauge bosons are predicted in many extensions to the
standard model (SM), such as left-right symmetric models and supersymmetric
grand unified theories~\cite{pgl_wprime, PhysRevD.11.566,PhysRevD.12.1502}.
The sensitivity to searches of new heavy bosons is usually explored
using a reference model
from Ref.~\cite{reference-model}, in
which the \Wprime\ is a copy of the $\PW$ boson with the same
left-handed fermionic couplings.  Interactions with the \SM gauge
bosons are excluded, as are interactions with other heavy gauge bosons
such as a ${\rm Z'}$. Thus, the \Wprime\ decay modes and branching
fractions are similar to those of the $\PW$ boson, with the
notable exception of the $\mathrm{t\bar{b}}$ channel, which opens for \Wprime\
masses beyond 180~\GeVcc. The leptonic branching fraction is
$\mathcal{B}(\Wprime \to \Pe \cPgn)=8.5\%$ for all masses
considered. 
In searches directly comparable to this one, the CDF and D0
experiments at the Fermilab Tevatron have excluded masses below
$1.1~\mathrm{TeV}/c^2$~\cite{cdf-limit-recent,d0-limit}. Here we
report the result of a \Wprime\ search using 36.1$\pm$4.0\pbinv of data
collected
between March and October 2010
in $pp$ collisions
at a center-of-mass energy of $\sqrt{s}=7$~TeV
with the Compact Muon Solenoid (CMS)
experiment at the CERN Large Hadron Collider (LHC).

A detailed description of the CMS detector can be found
elsewhere~\cite{cmsjinst:2008}.  We use a cylindrical coordinate
system about the beam axis in which $\theta$ is the polar angle with
respect to the counterclockwise beam direction, $\phi$ is the
azimuthal angle, and $\eta \equiv -\ln \tan (\theta /2)$. The
transverse energy is $E_{\rm T} \equiv E \sin \theta$, where $E$ is
defined as energy measured by the calorimeters.  The detectors used
for this analysis include the pixel and silicon strip trackers. They
provide coverage in the region $|\eta|<2.5$ and are immersed in a
$3.8$~T magnetic field to allow momentum determination of charged
particles. The electromagnetic and hadron calorimeters are used to
detect energy deposits from electrons in the range $|\eta|<2.5$ as
well as to provide an estimate of missing transverse energy due to escaping
particles. The electromagnetic calorimeter has an energy resolution of
better than 0.5\,\% for unconverted photons with transverse energies
above 100~GeV. The energy resolution is 3\,\% or better for the range
of electron energies relevant for this analysis.

Signal events would be characterized by the presence of a
high-energy electron and a large energy imbalance due to the
undetected escaping neutrino.  To acquire the data, we employ a
collection of single-electron triggers with several trigger
thresholds, which were changing frequently because of the evolving beam
conditions during 2010.  The bulk of the data were collected with a
trigger requiring an electron with $E_T^\mathrm{ele}>22$~GeV.  An
electron reconstruction algorithm optimized for high energy electrons
is used.  An electron is reconstructed as an energy deposit in the
electromagnetic calorimeter (referred to as a ``super-cluster'') with
a track pointing towards it. The electron track reconstruction is
based on a Gaussian sum filter algorithm~\cite{ptdr1} which takes into
account bremsstrahlung emissions along the electron trajectory.
Electron tracks pointing towards the transition region between the
barrel and endcap detectors in the electromagnetic calorimeter are
rejected.  Other requirements include matching criteria in $\eta$ and
$\phi$ between the super-cluster and the track, and the consistency of
the transverse shape of the energy deposit with that expected for an
electron. The electrons must have a transverse energy greater than
$30~\mathrm{GeV}$, and must be isolated in a cone of radius $\Delta R
\equiv \sqrt{\Delta \eta^2+\Delta\phi^2} < 0.3$ around the electron
candidate direction, both in the tracker and in the calorimeter. These
selections are designed to ensure high efficiency for electrons and a
high rejection of misreconstructed electrons from multi-jet
backgrounds.

To account for the energy imbalance due to the escaping neutrino, we
use a particle flow technique, which reconstructs a complete list of
particles in an event using all the available
information~\cite{PFT-09-001}.  Muons, electrons, photons, and charged
and neutral hadrons are reconstructed individually. We denote the
negative vector sum of the energy of all reconstructed particles in
the event projected on the transverse plane as \MET. It represents an
estimate of the vector sum of the transverse momentum of all escaping
neutral particles, such as neutrinos.

Since we are searching for a two-body decay which reconstructs to a high mass,
the energy of
the neutrino and electron are expected to be mostly balanced in the
transverse plane, both in direction and in magnitude. We therefore
require $0.4<E_T^\text{ele}/\MET<1.5$.  For the same reason, we
require that the angle between the electron and the \MET be close to
$\pi$ radians: $\Delta\phi_{e\MET}>2.5~\text{rad}$.

The primary discriminating variable is the transverse mass \MT. The
transverse mass is the equivalent of the invariant mass of a
four-vector computed with only the transverse components of those
four-vectors: $\MT = 
\sqrt{2 \cdot E_T^\text{ele} \cdot \MET \cdot (1 - \cos
  \Delta\phi_{e\MET}) }/c^2$.  As with a $\PW$, we expect the
\MT\ distribution of \Wprime\ events to exhibit a characteristic
Jacobian edge at the value of the mass of the decaying particle.
After applying these event selection criteria, the main background
consists of $\PW \to \Pe\cPgn$ decays in the tails of the \SM
W mass distribution.
Given the high transverse mass
of a potential signal, multi-jet production
constitutes a background when one jet is misreconstructed and the \MET
is a result of this misreconstruction.
Further contributions are due to $\PW \mathrm \to \tau \cPgn$ decays where the
tau-lepton subsequently decays to an electron and neutrinos. Since these
electrons result from several preceding decays, their energies are rather
low and do not spread much into the signal region.
Electrons from
semi-leptonic decays of \ttbar events may also constitute a background
primarily at low energies.
Small
contributions are caused by Drell-Yan ${\mathrm Z/\gamma}$ production followed
by decays into $\Pe^+ \Pe^-$ pairs where one electron is not detected,
diboson production and other QCD backgrounds such as ${\gamma}$+jet.
All backgrounds are summarized
in Tab.~\ref{tab:results_cutflow}.

\begin{figure*}[tbp]
\centering
\includegraphics[width=0.65\textwidth]{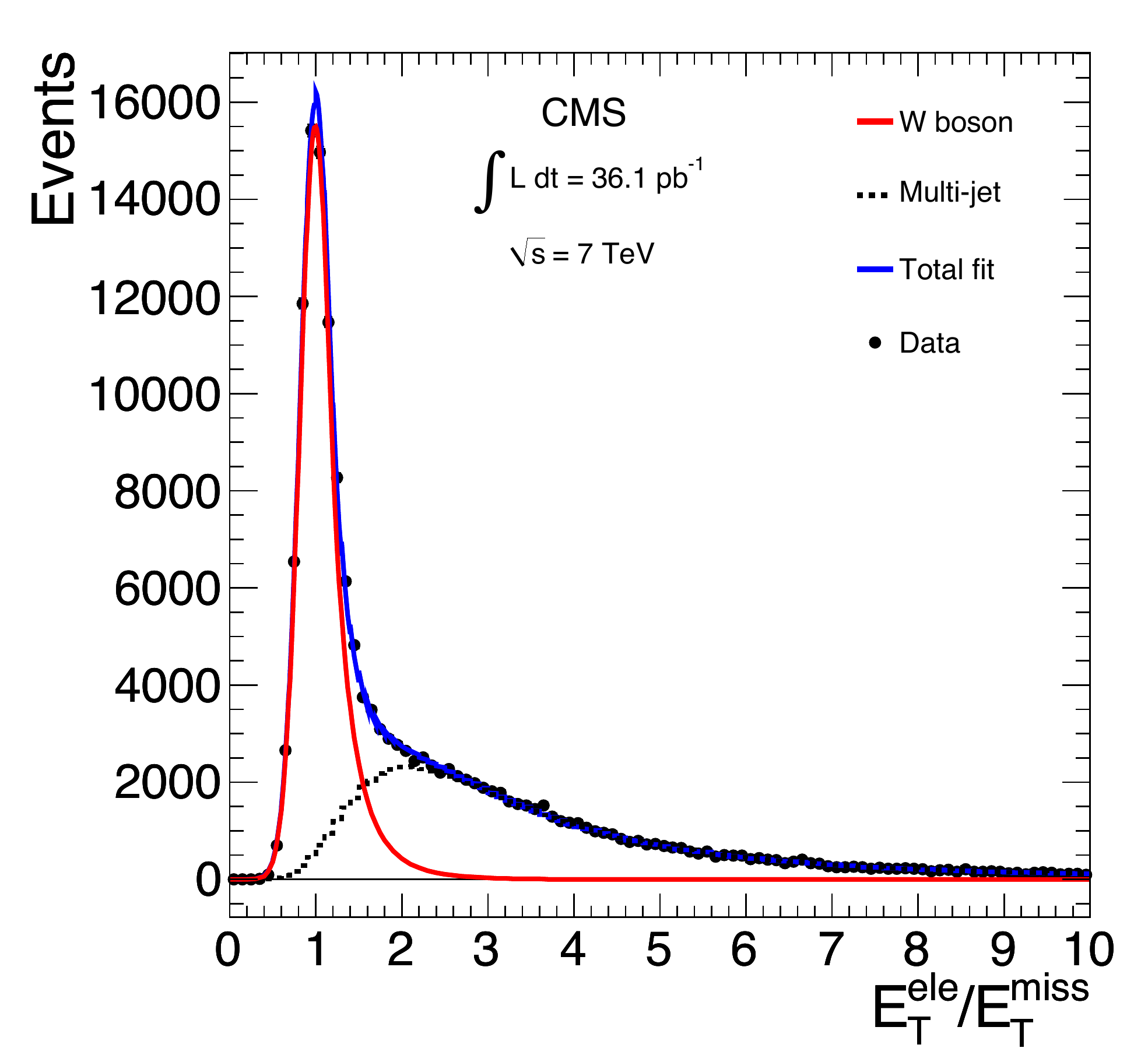}
\caption{
Distributions of the variable $E_T^\text{ele}/\MET$. The
  curves show the result of a simultaneous fit of the normalizations
  of multi-jet and $\PW$ background shapes to the data,
after subtraction of the
contribution of the other backgrounds.
}
\label{fig:background_W_qcd_fit_etOverMet}
\end{figure*}

Estimates of the \MT\ distribution for all backgrounds except for the
two
backgrounds, $\PW\to \Pe\cPgn$ and multi-jet events, are
obtained using Monte Carlo simulations, with a combination of
\textsc{ pythia v6.422}~\cite{pythia} and \textsc{madgraph}~\cite{madgraph}
event generators. The geometric and kinematic acceptances are
calculated using a \textsc{geant}-based simulation of the CMS
detector~\cite{geant4}. The \textsc{cteq6l1} parton distribution
functions are used to model the momentum distribution of the
initial-state partons~\cite{cteq6}.  For the signal sample, the {\sc
  pythia} event generator with its implementation of the
reference model~\cite{reference-model} is used at varying mass points up to
$M_{\Wprime}=2.0~\mathrm{TeV}/c^2$.
A mass-dependent $K$-factor of
about 1.3 approximating the next-to-next-to-leading-order (NNLO) cross
sections is applied~\cite{Hamberg:1991p2052,Hamberg:2002p2053},
varying between 1.26 (for $M_\Wprime=2~\TeVcc$) and 1.32 (for
$M_\Wprime=0.6~\TeVcc$).
After the final event selections, the product of geometric acceptance
and efficiency for the \Wprime\ signal is greater than 64\%,
independent of mass, for the range
$M_\Wprime=0.9-2.0~\mathrm{TeV}/c^2$.
The electron identification efficiency is measured
with the tag-and-probe method~\cite{EWK-10-002} in data and
simulation using a
clean sample of ${\mathrm Z \to \Pe^+ \Pe^-}$.
In this method one lepton candidate,
called the ``tag'', satisfies certain selection criteria while the
other lepton candidate, called the ``probe'', is required to pass
specific criteria which define the particular efficiency under study.
This method measured the electron identification efficiency to be
about 2\% lower in data than in Monte Carlo simulations and therefore
a correction factor was applied to the simulation.

The $\PW$ and multi-jet background estimates are derived from a
combination of experimental data and Monte Carlo simulations. For the
$\PW\to \Pe\cPgn$ backgrounds, we obtain an initial estimate of the
\MT\ shape from simulations using the {\sc pythia} event generator.
Differences in the \MET resolution between the data and simulations
are corrected using a technique based on the measurement of \MET in
${\mathrm Z\to \Pe^ + \Pe^-}$ events~\cite{EWK-10-002}.
The absolute normalizations of the multi-jet and W backgrounds
       are obtained using the $E_T^\text{ele}/\MET$
       distribution. A Crystal
       Ball~\cite{crystalball} function is used to describe the distribution for W bosons
       and an empirical shape is used for the multi-jet background,
       while for all other backgrounds the distributions and the
       normalisations are fixed to the standard model
       predictions. After subtraction of the latter backgrounds, a
       simultaneous fit to the $E_T^\text{ele}/\MET$ distribution in the data
       provides the multi-jet and W normalizations.
Figure~\ref{fig:background_W_qcd_fit_etOverMet} shows the
result of this fit.

\begin{table*}[tbp]
  \caption{Standard model background predictions and observed
    event counts, as a function of minimum \MT\ requirement, in units
    of \GeVcc. The
    errors include statistical and systematic uncertainties, but not the
    uncertainty on the luminosity normalization.
   }
  \begin{center}
    \begin{tabular}{|c|r@{$\pm\,$}rr@{$\pm\,$}lr@{$\pm\,$}lr@{$\pm\,$}lr@{$\pm\,$}lr@{$\pm\,$}l|} \hline
    Sample & \multicolumn{2}{c}{$>45$} &
    \multicolumn{2}{c}{$ > 200$} & \multicolumn{2}{c}{$ > 300$} &
    \multicolumn{2}{c}{$ > 400$} & \multicolumn{2}{c}{$ > 500$} &
    \multicolumn{2}{c|}{$ > 600$} \\ \hline
        $\PW \to \Pe \cPgn$
        & 75609 & 319 &  33.7 &  2.7 &   7.19 &  0.91 &   2.52 & 0.48 &  0.88 & 0.28 & 0.57 & 0.21 \\
        Multi-jet
        &  7083 & 3546   &   6.3 &  3.3 &   1.64 &  0.93 & 0.47 & 0.33 &  0.23 & 0.20 &  0.23 & 0.20 \\
        $\PW \mathrm \to \tau \cPgn$
        &  1083 & 80 &   1.1 &  0.3 &   0.21 &  0.19 &  \multicolumn{2}{c}{$<$0.13}   &  \multicolumn{2}{c}{$<$0.08} & \multicolumn{2}{c|}{$<$0.08} \\
        \ttbar
        &  60 & 23       &   4.1 &  1.7 &   0.64 &  0.29 & 0.15 & 0.09 &  0.03 & 0.03 &  0.01 & 0.02 \\
        Other bkg
        &  359 & 73      &   2.0 &  0.4 &   0.56 &  0.14 & 0.15 & 0.05 &  0.06 & 0.03 & 0.04 & 0.03  \\\hline
        Total bkg
        &  84194 & 3563  &  47.2 &  4.7 &  10.24 &  1.35 &  3.29 & 0.61 &  1.21 & 0.35 &  0.85 & 0.30  \\\hline
        Data     &    \multicolumn{2}{l}{84468} &  \multicolumn{2}{l}{38}   &   \multicolumn{2}{l}{8}   &   \multicolumn{2}{l}{2}    &   \multicolumn{2}{l}{0}   &   \multicolumn{2}{l|}{0} \\\hline
        \end{tabular}
  \end{center}
  \label{tab:results_cutflow}
\end{table*}

\begin{figure}[tbp]
  \begin{center}
    \includegraphics[width=0.65\columnwidth]{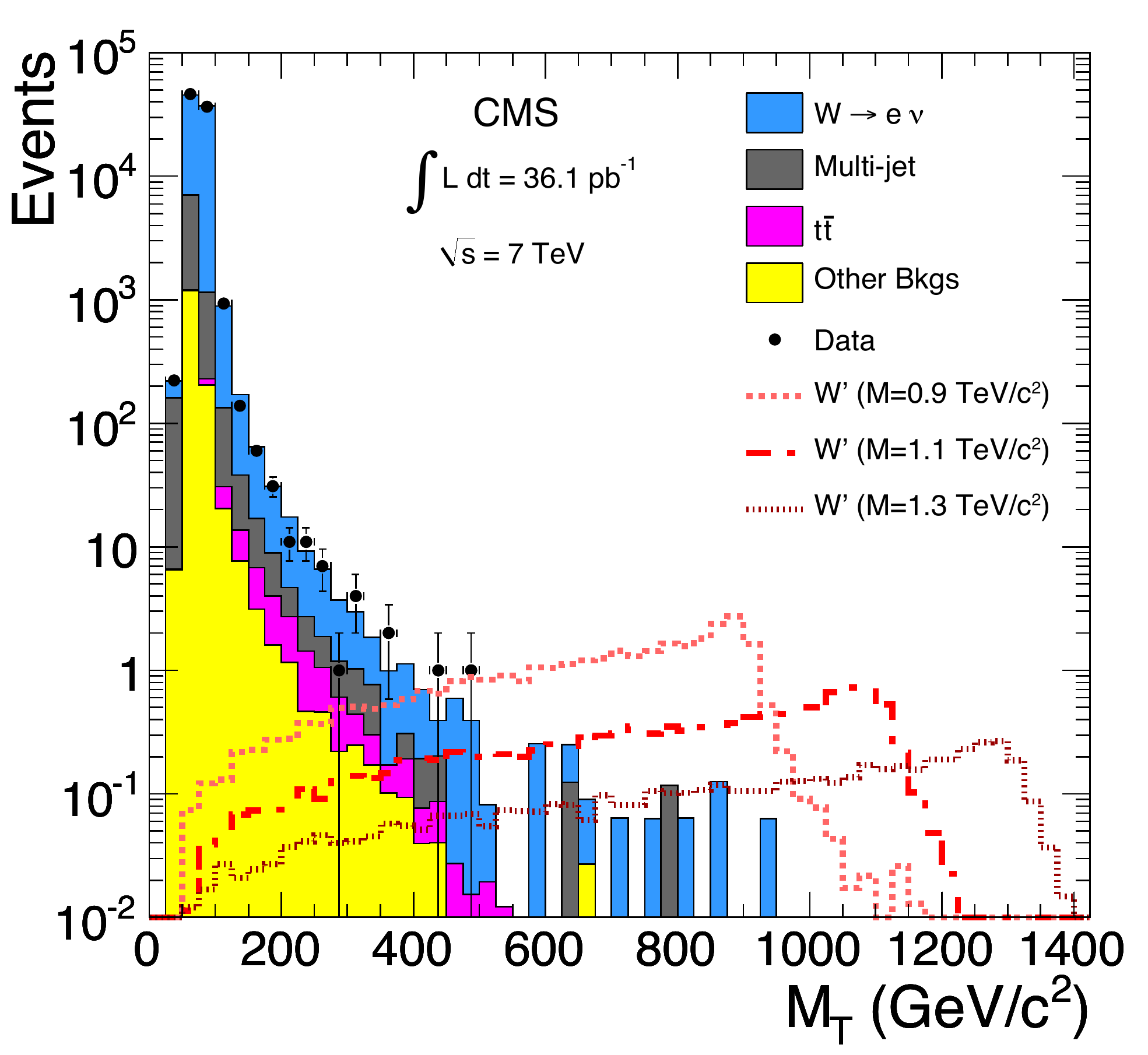}
  \end{center}
  \caption{Transverse mass distributions for all estimated standard model
    backgrounds and for CMS data. Here, other backgrounds also includes
    $\PW \mathrm \to \tau \cPgn$, in addition to the combined "other" backgrounds in
    Table~\ref{tab:results_cutflow}.
    The hashed lines represent the expected \MT\
    distributions for a \Wprime\ signal with three different \Wprime\
    masses.}
  \label{fig:results_mt_pf_log}
\end{figure}

Table~\ref{tab:results_cutflow}
shows the resulting number of predicted
background and observed events.  The background
estimate models the number of observed events well, including the low
transverse mass region ($\MT>45~\mathrm{GeV}/c^2$), which
contains most of the W-boson sample and
provides a
validation of the background model.
For \MT~$>$~400~\GeVcc the total expected background amounts to 3.29$\pm$0.61
events, out of which 75\% are caused by W bosons and 15\% by QCD.
No ${\PW \mathrm \to\tau\nu}$ event is expected and
a 68\% confidence limit (C.L.) upper limit is given in Table~\ref{tab:results_cutflow}.
Contributions caused by \ttbar are extremely small as well, of the same
order as
other backgrounds, which include Drell-Yan, dibosons and $\gamma$+jets.

In evaluating the systematic uncertainties, we consider estimates of
the acceptance and efficiency for reconstructing electrons (for the
\Wprime\ signal yield) as well as the uncertainty associated with the
background.
For those
backgrounds that are not derived from data, an uncertainty on
the absolute value of the integrated luminosity is included, taken as
11\%~\cite{EWK-10-004}, along with uncertainties on the cross sections
ranging from 5\% for diboson and $Z \to \ell^+ \ell^-$ to 39\% for
\ttbar~\cite{cmstopdil} using the CMS measurement for the latter.
With respect to electrons, reconstruction and identification
efficiency uncertainties of 1.9 and 1.5\%, respectively,
are included
for signal and backgrounds
as determined in an analysis of $Z\to \Pe^+ \Pe^-$ events.
For
the electron energy scale an
uncertainty of 1\% in the central section and 3\% for electrons in the
endcaps of the electromagnetic calorimeter~\cite{ecal-calibration} is included.
For the \MET resolution we assume an uncertainty of 10\%, applied as
an extra smearing to the
reconstructed
\MET in the simulation. The impact on the
number of events from all backgrounds is below 1\%.
A similar approach is used to evaluate the impact of the \MET scale.
A shift of 5\% is applied event by event to the \MET scale and the impact
on the event count for \MT~$>$~200~\GeVcc\ is found to be smaller than 10\%
for all backgrounds considered.
For the $\PW \to \Pe\cPgn$
background, the \MET and its associated uncertainty
are derived
using the hadronic recoil correction.
If we examine a region $\MT>200~\mathrm{GeV}/c^2$, the resulting
uncertainty on the predicted number of event counts is 8\% for the
main background $\PW \to \Pe \cPgn$ (dominated by the
uncertainties on the shape of the \MT\ distribution due to the energy
scale uncertainty) and 50\% for the sub-dominant multi-jet background
(studied by inverting the isolation requirement and analyzing the
multi-jet template in bins of \MT).  
The uncertainty on the number of signal events,
shown in Tab.~\ref{tab:fcnofmt},
is dominated by the
luminosity uncertainty (11\%) unlike the number of background events
which is dominated by the electron and \MET scale and resolution uncertainties (in total 28.7\%).
The limits reported in the same table are
relatively insensitive to systematic  uncertainties.

With all background estimates in hand, we examine the data for
evidence of non-\SM events. Figure~\ref{fig:results_mt_pf_log} shows
the CMS data overlaid with background expectations.  Since no excess
is observed in the data beyond the \SM background prediction, we set a
lower bound on the mass of the \Wprime\ boson in the reference model.
For each mass point, we choose a minimum \MT\ requirement that
provides the best {\it a priori} limit and use the \MT\ region above this
threshold as the search window.

\begin{figure*}[tp]
  \centering
    \includegraphics[width=0.45\textwidth]{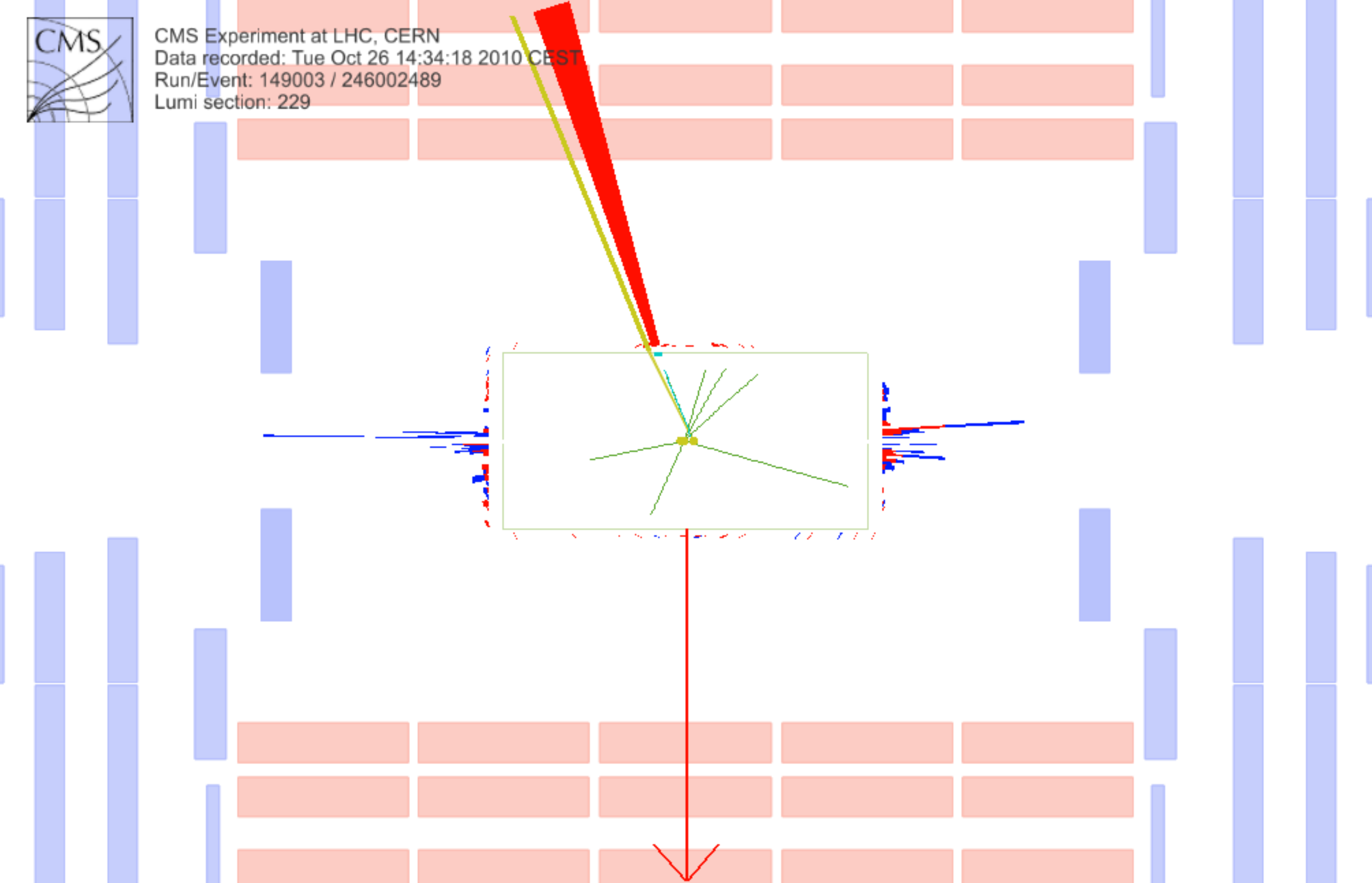}
    \includegraphics[width=0.45\textwidth]{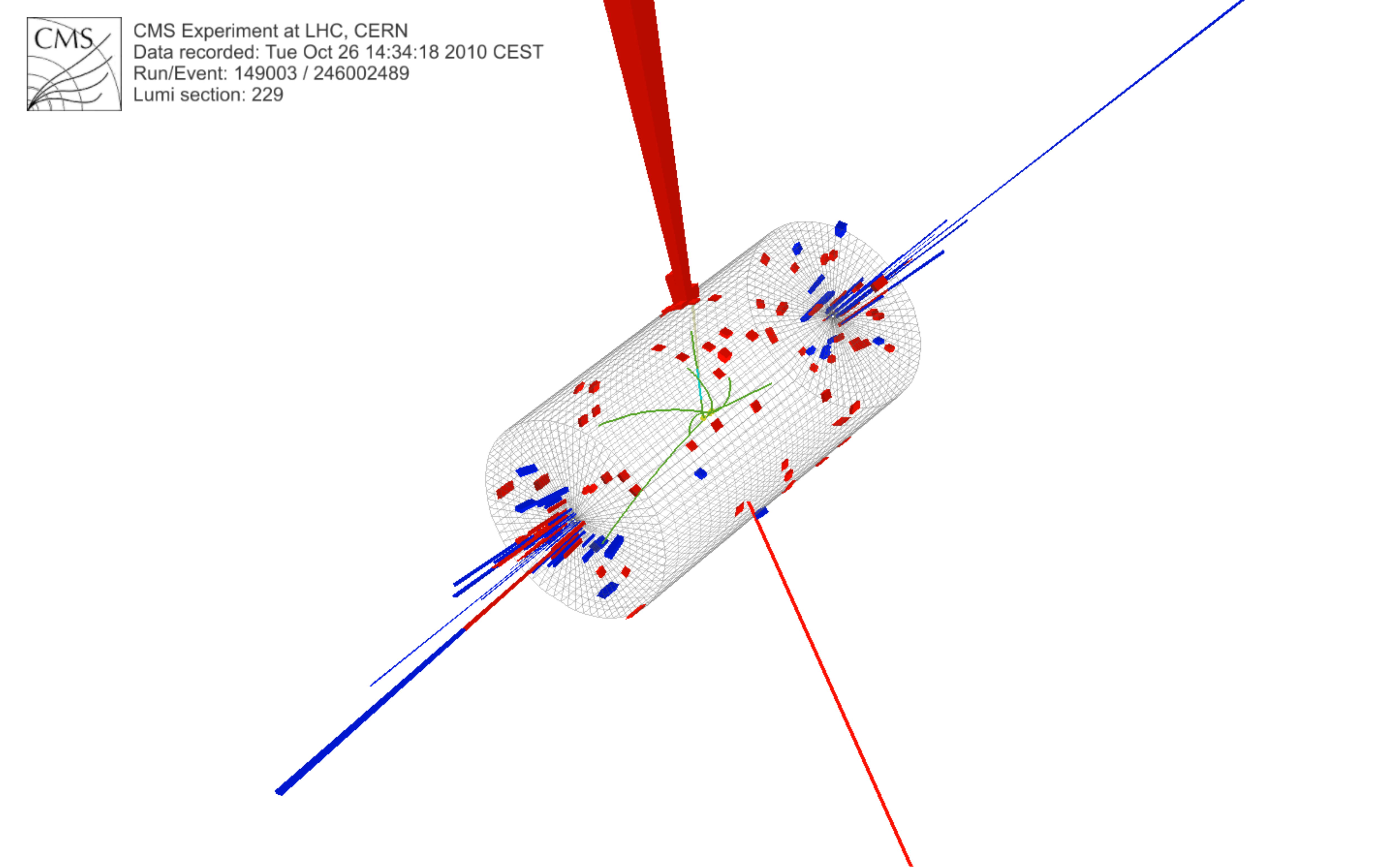}
    \caption{Displays of the highest \MT\ event. The projection on the
      left shows the envelope of the inner tracking detector along
      with the electromagnetic and hadron calorimeters, and part of
      the muon system.
      The 3D view on the right shows an enlarged
      view of the inner region.  Charged particle tracks as well as
      the deposited energy per calorimeter cell are displayed. The
      electron energy and \MET are shown in red, with the amount of
      energy represented graphically
      by the length of the bar.  }
  \label{fig:evd}
\end{figure*}

\begin{table*}[tp]
  \caption{Lower \MT\
    requirement as a function of \Wprime\ mass and expected and observed data counts.
    The entries $n_s,
    n_b$ and $n_d$ correspond to the expected signal and background
    counts and the observed data counts, respectively. The cross
    sections $\sigma_t, \sigma_e$ and $\sigma_o$ correspond to the
    theoretical \Wprime\ production cross section and  the expected
    and observed limits, respectively. The errors include all
    systematic uncertainties.}
  \centering
  \begin{tabular}{|rrccrrrr|}
  \hline
  $M_{\Wprime}$ & min \MT & $n_\text{s}$ & $n_\text{b}$ &
  $n_\text{d}$ & $\sigma_t$ & $\sigma_e$ & $\sigma_o$\\
  $(\mathrm{TeV}/c^2)$ & $(\mathrm{TeV}/c^2)$ & &&&$(\mathrm{pb})$ &$(\mathrm{pb})$ &$(\mathrm{pb})$ \\
  \hline
   0.6  &  0.400  &   129.38 $\pm$ 20.16  &  3.29 $\pm$ 0.61  &      2  &   8.290  &   0.379  &   0.289  \\
   0.7  &  0.500  &   60.77 $\pm$ 9.61  &  1.21 $\pm$ 0.35  &      0  &   4.264  &   0.314  &   0.215  \\
   0.8  &  0.500  &   39.54 $\pm$ 6.08  &  1.21 $\pm$ 0.35  &      0  &   2.426  &   0.274  &   0.188  \\
   0.9  &  0.500  &   25.24 $\pm$ 3.85  &  1.21 $\pm$ 0.35  &      0  &   1.389  &   0.246  &   0.168  \\
   1.0  &  0.500  &   16.10 $\pm$ 2.45  &  1.21 $\pm$ 0.35  &      0  &   0.838  &   0.232  &   0.159  \\
   1.1  &  0.500  &   10.06 $\pm$ 1.53  &  1.21 $\pm$ 0.35  &      0  &   0.516  &   0.229  &   0.157  \\
   1.2  &  0.650  &   6.02 $\pm$ 0.92  &  0.60 $\pm$ 0.24  &      0  &   0.334  &   0.215  &   0.170  \\
   1.3  &  0.675  &   3.92 $\pm$ 0.60  &  0.51 $\pm$ 0.21  &      0  &   0.215  &   0.207  &   0.168  \\
   1.4  &  0.675  &   2.52 $\pm$ 0.38  &  0.51 $\pm$ 0.21  &      0  &   0.136  &   0.203  &   0.164  \\
   1.5  &  0.675  &   1.89 $\pm$ 0.29  &  0.51 $\pm$ 0.21  &      0  &   0.099  &   0.196  &   0.159  \\
   2.0  &  0.675  &   0.27 $\pm$ 0.04  &  0.51 $\pm$ 0.21  &      0  &   0.014  &   0.206  &   0.167  \\ \hline
 \end{tabular}
 \label{tab:fcnofmt}
\end{table*}
The resultant minimum \MT\ requirement ranges from
$400$ to $675~\mathrm{GeV}/c^2$ across the \Wprime\ mass range and is shown
in Tab.~\ref{tab:fcnofmt}
along with the corresponding number of potential signal and background events.
The errors include all systematic uncertainties.
The number of events found in data is also shown.  The highest
transverse mass event we observe has $\MT=493~\mathrm{GeV}/c^2$ and is
shown in Fig.~\ref{fig:evd}.

\begin{figure*}[tbp]
  \begin{center}
    \includegraphics[width=0.65\textwidth]{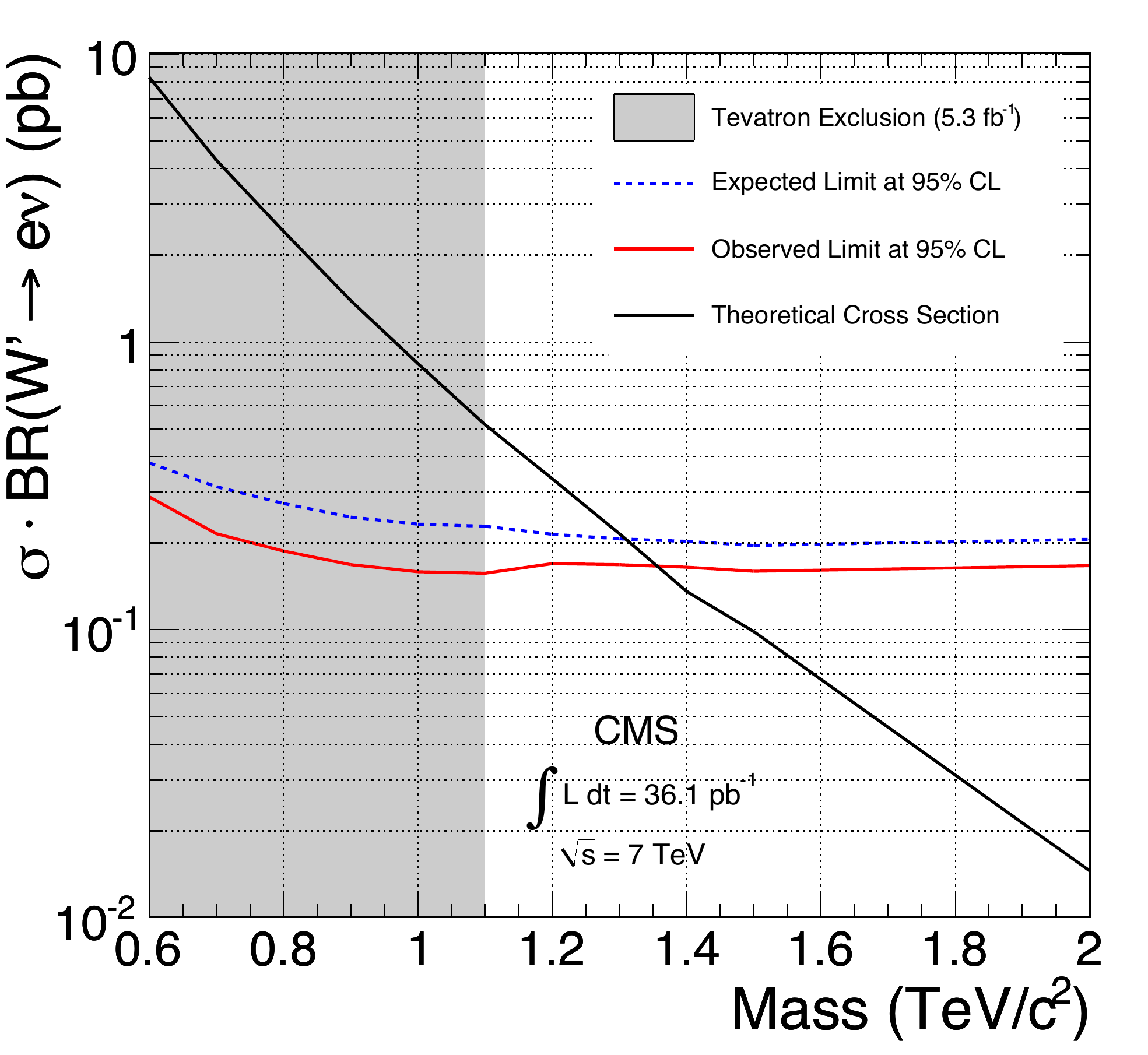}
    \caption{Limit using a Bayesian technique with a counting
      experiment in the search window, for the reference model. The
      intersection of the cross section limit curve and
      the theoretical cross section yields a lower limit of
      $M_{\Wprime}>1.36~\mathrm{TeV}/c^2$ at 95\% C.L. for the assumed
      $\sigma \mathcal{\times B}(\Wprime\to \Pe \nu)$.}
    \label{fig:limit_cls_MT500}
  \end{center}
\end{figure*}

We use a Bayesian technique to determine an upper limit on the cross
section as a function of the \Wprime\ boson mass with a
C.L.  of 95$\%$, following the method described
in~\cite{Fermilab-TM-2104}.  We assume a flat prior probability distribution
for the cross
section. To incorporate the systematic uncertainties described above,
we treat them as nuisance parameters and use a log-normal
distribution to integrate over these parameters.
Figure~\ref{fig:limit_cls_MT500} shows both the expected and the
observed limit. We exclude the existence of \Wprime\ bosons with
standard model-like couplings and branching fractions with masses
up to 1.36~$\mathrm{TeV}/c^{2}$ at 95$\%$ C.L.

In summary, we have performed a search for a heavy gauge boson
\Wprime\ in the decay channel with an electron and large transverse
energy imbalance. We observe no excess over the background.
A new W boson-like gauge particle with standard-model-like couplings and
branching fractions up to a mass of $1.36~\mathrm{TeV}/c^2$ is
excluded by the data, the most stringent limit to date.

We wish to congratulate our colleagues in the CERN accelerator
departments for the excellent performance of the LHC machine. We thank
the technical and administrative staff at CERN and other CMS
institutes, and acknowledge support from: FMSR (Austria); FNRS and FWO
(Belgium); CNPq, CAPES, FAPERJ, and FAPESP (Brazil); MES (Bulgaria);
CERN; CAS, MoST, and NSFC (China); COLCIENCIAS (Colombia); MSES
(Croatia); RPF (Cyprus); Academy of Sciences and NICPB (Estonia);
Academy of Finland, ME, and HIP (Finland); CEA and CNRS/IN2P3
(France); BMBF, DFG, and HGF (Germany); GSRT (Greece); OTKA and NKTH
(Hungary); DAE and DST (India); IPM (Iran); SFI (Ireland); INFN
(Italy); NRF and WCU (Korea); LAS (Lithuania); CINVESTAV, CONACYT,
SEP, and UASLP-FAI (Mexico); PAEC (Pakistan); SCSR (Poland); FCT
(Portugal); JINR (Armenia, Belarus, Georgia, Ukraine, Uzbekistan); MST
and MAE (Russia); MSTD (Serbia); MICINN and CPAN (Spain); Swiss
Funding Agencies (Switzerland); NSC (Taipei); TUBITAK and TAEK
(Turkey); STFC (United Kingdom); DOE and NSF (USA).
\bibliography{auto_generated}   

\cleardoublepage\appendix\section{The CMS Collaboration \label{app:collab}}\begin{sloppypar}\hyphenpenalty=5000\widowpenalty=500\clubpenalty=5000\input{EXO-10-014-authorlist.tex}\end{sloppypar}
\end{document}

%% file: ptdr-definitions.tex
%
%
%

\providecommand {\etal}{\mbox{et al.}\xspace} 
\providecommand {\ie}{\mbox{i.e.}\xspace}     
\providecommand {\eg}{\mbox{e.g.}\xspace}     
\providecommand {\etc}{\mbox{etc.}\xspace}     
\providecommand {\vs}{\mbox{\sl vs.}\xspace}      
\providecommand {\mdash}{\ensuremath{\mathrm{-}}} 

\providecommand {\Lone}{Level-1\xspace} 
\providecommand {\Ltwo}{Level-2\xspace}
\providecommand {\Lthree}{Level-3\xspace}

\providecommand{\ACERMC} {\textsc{AcerMC}\xspace}
\providecommand{\ALPGEN} {{\textsc{alpgen}}\xspace}
\providecommand{\CHARYBDIS} {{\textsc{charybdis}}\xspace}
\providecommand{\CMKIN} {\textsc{cmkin}\xspace}
\providecommand{\CMSIM} {{\textsc{cmsim}}\xspace}
\providecommand{\CMSSW} {{\textsc{cmssw}}\xspace}
\providecommand{\COBRA} {{\textsc{cobra}}\xspace}
\providecommand{\COCOA} {{\textsc{cocoa}}\xspace}
\providecommand{\COMPHEP} {\textsc{CompHEP}\xspace}
\providecommand{\EVTGEN} {{\textsc{evtgen}}\xspace}
\providecommand{\FAMOS} {{\textsc{famos}}\xspace}
\providecommand{\GARCON} {\textsc{garcon}\xspace}
\providecommand{\GARFIELD} {{\textsc{garfield}}\xspace}
\providecommand{\GEANE} {{\textsc{geane}}\xspace}
\providecommand{\GEANTfour} {{\textsc{geant4}}\xspace}
\providecommand{\GEANTthree} {{\textsc{geant3}}\xspace}
\providecommand{\GEANT} {{\textsc{geant}}\xspace}
\providecommand{\HDECAY} {\textsc{hdecay}\xspace}
\providecommand{\HERWIG} {{\textsc{herwig}}\xspace}
\providecommand{\HIGLU} {{\textsc{higlu}}\xspace}
\providecommand{\HIJING} {{\textsc{hijing}}\xspace}
\providecommand{\IGUANA} {\textsc{iguana}\xspace}
\providecommand{\ISAJET} {{\textsc{isajet}}\xspace}
\providecommand{\ISAPYTHIA} {{\textsc{isapythia}}\xspace}
\providecommand{\ISASUGRA} {{\textsc{isasugra}}\xspace}
\providecommand{\ISASUSY} {{\textsc{isasusy}}\xspace}
\providecommand{\ISAWIG} {{\textsc{isawig}}\xspace}
\providecommand{\MADGRAPH} {\textsc{MadGraph}\xspace}
\providecommand{\MCATNLO} {\textsc{mc@nlo}\xspace}
\providecommand{\MCFM} {\textsc{mcfm}\xspace}
\providecommand{\MILLEPEDE} {{\textsc{millepede}}\xspace}
\providecommand{\ORCA} {{\textsc{orca}}\xspace}
\providecommand{\OSCAR} {{\textsc{oscar}}\xspace}
\providecommand{\PHOTOS} {\textsc{photos}\xspace}
\providecommand{\PROSPINO} {\textsc{prospino}\xspace}
\providecommand{\PYTHIA} {{\textsc{pythia}}\xspace}
\providecommand{\SHERPA} {{\textsc{sherpa}}\xspace}
\providecommand{\TAUOLA} {\textsc{tauola}\xspace}
\providecommand{\TOPREX} {\textsc{TopReX}\xspace}
\providecommand{\XDAQ} {{\textsc{xdaq}}\xspace}

\providecommand {\DZERO}{D\O\xspace}     


\providecommand{\de}{\ensuremath{^\circ}}
\providecommand{\ten}[1]{\ensuremath{\times \text{10}^\text{#1}}}
\providecommand{\unit}[1]{\ensuremath{\text{\,#1}}\xspace}
\providecommand{\mum}{\ensuremath{\,\mu\text{m}}\xspace}
\providecommand{\micron}{\ensuremath{\,\mu\text{m}}\xspace}
\providecommand{\cm}{\ensuremath{\,\text{cm}}\xspace}
\providecommand{\mm}{\ensuremath{\,\text{mm}}\xspace}
\providecommand{\mus}{\ensuremath{\,\mu\text{s}}\xspace}
\providecommand{\keV}{\ensuremath{\,\text{ke\hspace{-.08em}V}}\xspace}
\providecommand{\MeV}{\ensuremath{\,\text{Me\hspace{-.08em}V}}\xspace}
\providecommand{\GeV}{\ensuremath{\,\text{Ge\hspace{-.08em}V}}\xspace}
\providecommand{\gev}{\GeV}
\providecommand{\TeV}{\ensuremath{\,\text{Te\hspace{-.08em}V}}\xspace}
\providecommand{\PeV}{\ensuremath{\,\text{Pe\hspace{-.08em}V}}\xspace}
\providecommand{\keVc}{\ensuremath{{\,\text{ke\hspace{-.08em}V\hspace{-0.16em}/\hspace{-0.08em}}c}}\xspace}
\providecommand{\MeVc}{\ensuremath{{\,\text{Me\hspace{-.08em}V\hspace{-0.16em}/\hspace{-0.08em}}c}}\xspace}
\providecommand{\GeVc}{\ensuremath{{\,\text{Ge\hspace{-.08em}V\hspace{-0.16em}/\hspace{-0.08em}}c}}\xspace}
\providecommand{\TeVc}{\ensuremath{{\,\text{Te\hspace{-.08em}V\hspace{-0.16em}/\hspace{-0.08em}}c}}\xspace}
\providecommand{\keVcc}{\ensuremath{{\,\text{ke\hspace{-.08em}V\hspace{-0.16em}/\hspace{-0.08em}}c^\text{2}}}\xspace}
\providecommand{\MeVcc}{\ensuremath{{\,\text{Me\hspace{-.08em}V\hspace{-0.16em}/\hspace{-0.08em}}c^\text{2}}}\xspace}
\providecommand{\GeVcc}{\ensuremath{{\,\text{Ge\hspace{-.08em}V\hspace{-0.16em}/\hspace{-0.08em}}c^\text{2}}}\xspace}
\providecommand{\TeVcc}{\ensuremath{{\,\text{Te\hspace{-.08em}V\hspace{-0.16em}/\hspace{-0.08em}}c^\text{2}}}\xspace}

\providecommand{\pbinv} {\mbox{\ensuremath{\,\text{pb}^\text{$-$1}}}\xspace}
\providecommand{\fbinv} {\mbox{\ensuremath{\,\text{fb}^\text{$-$1}}}\xspace}
\providecommand{\nbinv} {\mbox{\ensuremath{\,\text{nb}^\text{$-$1}}}\xspace}
\providecommand{\percms}{\ensuremath{\,\text{cm}^\text{$-$2}\,\text{s}^\text{$-$1}}\xspace}
\providecommand{\lumi}{\ensuremath{\mathcal{L}}\xspace}
\providecommand{\Lumi}{\ensuremath{\mathcal{L}}\xspace}
%
\providecommand{\LvLow}  {\ensuremath{\mathcal{L}=\text{10}^\text{32}\,\text{cm}^\text{$-$2}\,\text{s}^\text{$-$1}}\xspace}
\providecommand{\LLow}   {\ensuremath{\mathcal{L}=\text{10}^\text{33}\,\text{cm}^\text{$-$2}\,\text{s}^\text{$-$1}}\xspace}
\providecommand{\lowlumi}{\ensuremath{\mathcal{L}=\text{2}\times \text{10}^\text{33}\,\text{cm}^\text{$-$2}\,\text{s}^\text{$-$1}}\xspace}
\providecommand{\LMed}   {\ensuremath{\mathcal{L}=\text{2}\times \text{10}^\text{33}\,\text{cm}^\text{$-$2}\,\text{s}^\text{$-$1}}\xspace}
\providecommand{\LHigh}  {\ensuremath{\mathcal{L}=\text{10}^\text{34}\,\text{cm}^\text{$-$2}\,\text{s}^\text{$-$1}}\xspace}
\providecommand{\hilumi} {\ensuremath{\mathcal{L}=\text{10}^\text{34}\,\text{cm}^\text{$-$2}\,\text{s}^\text{$-$1}}\xspace}


\providecommand{\PT}{\ensuremath{p_{\mathrm{T}}}\xspace}
\providecommand{\pt}{\ensuremath{p_{\mathrm{T}}}\xspace}
\providecommand{\ET}{\ensuremath{E_{\mathrm{T}}}\xspace}
\providecommand{\HT}{\ensuremath{H_{\mathrm{T}}}\xspace}
\providecommand{\et}{\ensuremath{E_{\mathrm{T}}}\xspace}
\providecommand{\Em}{\ensuremath{E\hspace{-0.6em}/}\xspace}
\providecommand{\Pm}{\ensuremath{p\hspace{-0.5em}/}\xspace}
\providecommand{\PTm}{\ensuremath{{p}_\mathrm{T}\hspace{-1.02em}/}\xspace}
\providecommand{\PTslash}{\ensuremath{{p}_\mathrm{T}\hspace{-1.02em}/}\xspace}
\providecommand{\ETm}{\ensuremath{E_{\mathrm{T}}^{\text{miss}}}\xspace}
\providecommand{\ETslash}{\ensuremath{E_{\mathrm{T}}\hspace{-1.1em}/}\xspace}
\providecommand{\MET}{\ensuremath{E_{\mathrm{T}}^{\text{miss}}}\xspace}
\providecommand{\ETmiss}{\ensuremath{E_{\mathrm{T}}^{\text{miss}}}\xspace}
\providecommand{\VEtmiss}{\ensuremath{{\vec E}_{\mathrm{T}}^{\text{miss}}}\xspace}

\providecommand{\dd}[2]{\ensuremath{\frac{\mathrm{d} #1}{\mathrm{d} #2}}}

\ifthenelse{\boolean{cms@italic}}{\newcommand{\cmsSymbolFace}{\relax}}{\newcommand{\cmsSymbolFace}{\mathrm}}

\providecommand{\zp}{\ensuremath{\cmsSymbolFace{Z}^\prime}\xspace}
\providecommand{\JPsi}{\ensuremath{\cmsSymbolFace{J}\hspace{-.08em}/\hspace{-.14em}\psi}\xspace}
\providecommand{\Z}{\ensuremath{\cmsSymbolFace{Z}}\xspace}
\providecommand{\ttbar}{\ensuremath{\cmsSymbolFace{t}\overline{\cmsSymbolFace{t}}}\xspace}

\newcommand{\cPgn}{\ensuremath{\nu}}
\newcommand{\cPJgy}{\JPsi}
\newcommand{\cPZ}{\Z}
\newcommand{\cPZpr}{\zp}


\providecommand{\AFB}{\ensuremath{A_\text{FB}}\xspace}
\providecommand{\wangle}{\ensuremath{\sin^{2}\theta_{\text{eff}}^\text{lept}(M^2_\Z)}\xspace}
\providecommand{\stat}{\ensuremath{\,\text{(stat.)}}\xspace}
\providecommand{\syst}{\ensuremath{\,\text{(syst.)}}\xspace}
\providecommand{\kt}{\ensuremath{k_{\mathrm{T}}}\xspace}

\providecommand{\BC}{\ensuremath{\mathrm{B_{c}}}\xspace}
\providecommand{\bbarc}{\ensuremath{\mathrm{\overline{b}c}}\xspace}
\providecommand{\bbbar}{\ensuremath{\mathrm{b\overline{b}}}\xspace}
\providecommand{\ccbar}{\ensuremath{\mathrm{c\overline{c}}}\xspace}
\providecommand{\bspsiphi}{\ensuremath{\mathrm{B_s} \to \JPsi\, \phi}\xspace}
\providecommand{\EE}{\ensuremath{\mathrm{e^+e^-}}\xspace}
\providecommand{\MM}{\ensuremath{\mu^+\mu^-}\xspace}
\providecommand{\TT}{\ensuremath{\tau^+\tau^-}\xspace}

\providecommand{\HGG}{\ensuremath{\mathrm{H}\to\gamma\gamma}}
\providecommand{\GAMJET}{\ensuremath{\gamma + \text{jet}}}
\providecommand{\PPTOJETS}{\ensuremath{\mathrm{pp}\to\text{jets}}}
\providecommand{\PPTOGG}{\ensuremath{\mathrm{pp}\to\gamma\gamma}}
\providecommand{\PPTOGAMJET}{\ensuremath{\mathrm{pp}\to\gamma + \mathrm{jet}}}
\providecommand{\MH}{\ensuremath{M_{\mathrm{H}}}}
\providecommand{\RNINE}{\ensuremath{R_\mathrm{9}}}
\providecommand{\DR}{\ensuremath{\Delta R}}

%

\providecommand{\ga}{\ensuremath{\gtrsim}}
\providecommand{\la}{\ensuremath{\lesssim}}
\providecommand{\swsq}{\ensuremath{\sin^2\theta_\mathrm{W}}\xspace}
\providecommand{\cwsq}{\ensuremath{\cos^2\theta_\mathrm{W}}\xspace}
\providecommand{\tanb}{\ensuremath{\tan\beta}\xspace}
\providecommand{\tanbsq}{\ensuremath{\tan^{2}\beta}\xspace}
\providecommand{\sidb}{\ensuremath{\sin 2\beta}\xspace}
\providecommand{\alpS}{\ensuremath{\alpha_S}\xspace}
\providecommand{\alpt}{\ensuremath{\tilde{\alpha}}\xspace}

\providecommand{\QL}{\ensuremath{\mathrm{Q}_\mathrm{L}}\xspace}
\providecommand{\sQ}{\ensuremath{\tilde{\mathrm{Q}}}\xspace}
\providecommand{\sQL}{\ensuremath{\tilde{\mathrm{Q}}_\mathrm{L}}\xspace}
\providecommand{\ULC}{\ensuremath{\mathrm{U}_\mathrm{L}^\mathrm{C}}\xspace}
\providecommand{\sUC}{\ensuremath{\tilde{\mathrm{U}}^\mathrm{C}}\xspace}
\providecommand{\sULC}{\ensuremath{\tilde{\mathrm{U}}_\mathrm{L}^\mathrm{C}}\xspace}
\providecommand{\DLC}{\ensuremath{\mathrm{D}_\mathrm{L}^\mathrm{C}}\xspace}
\providecommand{\sDC}{\ensuremath{\tilde{\mathrm{D}}^\mathrm{C}}\xspace}
\providecommand{\sDLC}{\ensuremath{\tilde{\mathrm{D}}_\mathrm{L}^\mathrm{C}}\xspace}
\providecommand{\LL}{\ensuremath{\mathrm{L}_\mathrm{L}}\xspace}
\providecommand{\sL}{\ensuremath{\tilde{\mathrm{L}}}\xspace}
\providecommand{\sLL}{\ensuremath{\tilde{\mathrm{L}}_\mathrm{L}}\xspace}
\providecommand{\ELC}{\ensuremath{\mathrm{E}_\mathrm{L}^\mathrm{C}}\xspace}
\providecommand{\sEC}{\ensuremath{\tilde{\mathrm{E}}^\mathrm{C}}\xspace}
\providecommand{\sELC}{\ensuremath{\tilde{\mathrm{E}}_\mathrm{L}^\mathrm{C}}\xspace}
\providecommand{\sEL}{\ensuremath{\tilde{\mathrm{E}}_\mathrm{L}}\xspace}
\providecommand{\sER}{\ensuremath{\tilde{\mathrm{E}}_\mathrm{R}}\xspace}
\providecommand{\sFer}{\ensuremath{\tilde{\mathrm{f}}}\xspace}
\providecommand{\sQua}{\ensuremath{\tilde{\mathrm{q}}}\xspace}
\providecommand{\sUp}{\ensuremath{\tilde{\mathrm{u}}}\xspace}
\providecommand{\suL}{\ensuremath{\tilde{\mathrm{u}}_\mathrm{L}}\xspace}
\providecommand{\suR}{\ensuremath{\tilde{\mathrm{u}}_\mathrm{R}}\xspace}
\providecommand{\sDw}{\ensuremath{\tilde{\mathrm{d}}}\xspace}
\providecommand{\sdL}{\ensuremath{\tilde{\mathrm{d}}_\mathrm{L}}\xspace}
\providecommand{\sdR}{\ensuremath{\tilde{\mathrm{d}}_\mathrm{R}}\xspace}
\providecommand{\sTop}{\ensuremath{\tilde{\mathrm{t}}}\xspace}
\providecommand{\stL}{\ensuremath{\tilde{\mathrm{t}}_\mathrm{L}}\xspace}
\providecommand{\stR}{\ensuremath{\tilde{\mathrm{t}}_\mathrm{R}}\xspace}
\providecommand{\stone}{\ensuremath{\tilde{\mathrm{t}}_1}\xspace}
\providecommand{\sttwo}{\ensuremath{\tilde{\mathrm{t}}_2}\xspace}
\providecommand{\sBot}{\ensuremath{\tilde{\mathrm{b}}}\xspace}
\providecommand{\sbL}{\ensuremath{\tilde{\mathrm{b}}_\mathrm{L}}\xspace}
\providecommand{\sbR}{\ensuremath{\tilde{\mathrm{b}}_\mathrm{R}}\xspace}
\providecommand{\sbone}{\ensuremath{\tilde{\mathrm{b}}_1}\xspace}
\providecommand{\sbtwo}{\ensuremath{\tilde{\mathrm{b}}_2}\xspace}
\providecommand{\sLep}{\ensuremath{\tilde{\mathrm{l}}}\xspace}
\providecommand{\sLepC}{\ensuremath{\tilde{\mathrm{l}}^\mathrm{C}}\xspace}
\providecommand{\sEl}{\ensuremath{\tilde{\mathrm{e}}}\xspace}
\providecommand{\sElC}{\ensuremath{\tilde{\mathrm{e}}^\mathrm{C}}\xspace}
\providecommand{\seL}{\ensuremath{\tilde{\mathrm{e}}_\mathrm{L}}\xspace}
\providecommand{\seR}{\ensuremath{\tilde{\mathrm{e}}_\mathrm{R}}\xspace}
\providecommand{\snL}{\ensuremath{\tilde{\nu}_L}\xspace}
\providecommand{\sMu}{\ensuremath{\tilde{\mu}}\xspace}
\providecommand{\sNu}{\ensuremath{\tilde{\nu}}\xspace}
\providecommand{\sTau}{\ensuremath{\tilde{\tau}}\xspace}
\providecommand{\Glu}{\ensuremath{\mathrm{g}}\xspace}
\providecommand{\sGlu}{\ensuremath{\tilde{\mathrm{g}}}\xspace}
\providecommand{\Wpm}{\ensuremath{\mathrm{W}^{\pm}}\xspace}
\providecommand{\sWpm}{\ensuremath{\tilde{\mathrm{W}}^{\pm}}\xspace}
\providecommand{\Wz}{\ensuremath{\mathrm{W}^{0}}\xspace}
\providecommand{\sWz}{\ensuremath{\tilde{\mathrm{W}}^{0}}\xspace}
\providecommand{\sWino}{\ensuremath{\tilde{\mathrm{W}}}\xspace}
\providecommand{\Bz}{\ensuremath{\mathrm{B}^{0}}\xspace}
\providecommand{\sBz}{\ensuremath{\tilde{\mathrm{B}}^{0}}\xspace}
\providecommand{\sBino}{\ensuremath{\tilde{\mathrm{B}}}\xspace}
\providecommand{\Zz}{\ensuremath{\mathrm{Z}^{0}}\xspace}
\providecommand{\sZino}{\ensuremath{\tilde{\mathrm{Z}}^{0}}\xspace}
\providecommand{\sGam}{\ensuremath{\tilde{\gamma}}\xspace}
\providecommand{\chiz}{\ensuremath{\tilde{\chi}^{0}}\xspace}
\providecommand{\chip}{\ensuremath{\tilde{\chi}^{+}}\xspace}
\providecommand{\chim}{\ensuremath{\tilde{\chi}^{-}}\xspace}
\providecommand{\chipm}{\ensuremath{\tilde{\chi}^{\pm}}\xspace}
\providecommand{\Hone}{\ensuremath{\mathrm{H}_\mathrm{d}}\xspace}
\providecommand{\sHone}{\ensuremath{\tilde{\mathrm{H}}_\mathrm{d}}\xspace}
\providecommand{\Htwo}{\ensuremath{\mathrm{H}_\mathrm{u}}\xspace}
\providecommand{\sHtwo}{\ensuremath{\tilde{\mathrm{H}}_\mathrm{u}}\xspace}
\providecommand{\sHig}{\ensuremath{\tilde{\mathrm{H}}}\xspace}
\providecommand{\sHa}{\ensuremath{\tilde{\mathrm{H}}_\mathrm{a}}\xspace}
\providecommand{\sHb}{\ensuremath{\tilde{\mathrm{H}}_\mathrm{b}}\xspace}
\providecommand{\sHpm}{\ensuremath{\tilde{\mathrm{H}}^{\pm}}\xspace}
\providecommand{\hz}{\ensuremath{\mathrm{h}^{0}}\xspace}
\providecommand{\Hz}{\ensuremath{\mathrm{H}^{0}}\xspace}
\providecommand{\Az}{\ensuremath{\mathrm{A}^{0}}\xspace}
\providecommand{\Hpm}{\ensuremath{\mathrm{H}^{\pm}}\xspace}
\providecommand{\sGra}{\ensuremath{\tilde{\mathrm{G}}}\xspace}
\providecommand{\mtil}{\ensuremath{\tilde{m}}\xspace}
\providecommand{\rpv}{\ensuremath{\rlap{\kern.2em/}R}\xspace}
\providecommand{\LLE}{\ensuremath{LL\bar{E}}\xspace}
\providecommand{\LQD}{\ensuremath{LQ\bar{D}}\xspace}
\providecommand{\UDD}{\ensuremath{\overline{UDD}}\xspace}
\providecommand{\Lam}{\ensuremath{\lambda}\xspace}
\providecommand{\Lamp}{\ensuremath{\lambda'}\xspace}
\providecommand{\Lampp}{\ensuremath{\lambda''}\xspace}
\providecommand{\spinbd}[2]{\ensuremath{\bar{#1}_{\dot{#2}}}\xspace}

\providecommand{\MD}{\ensuremath{{M_\mathrm{D}}}\xspace}
\providecommand{\Mpl}{\ensuremath{{M_\mathrm{Pl}}}\xspace}
\providecommand{\Rinv} {\ensuremath{{R}^{-1}}\xspace} 

%% file: EXO-10-014-authorlist.tex
\textbf{Yerevan Physics Institute,  Yerevan,  Armenia}\\*[0pt]
V.~Khachatryan, A.M.~Sirunyan, A.~Tumasyan
\vskip\cmsinstskip
\textbf{Institut f\"{u}r Hochenergiephysik der OeAW,  Wien,  Austria}\\*[0pt]
W.~Adam, T.~Bergauer, M.~Dragicevic, J.~Er\"{o}, C.~Fabjan, M.~Friedl, R.~Fr\"{u}hwirth, V.M.~Ghete, J.~Hammer\cmsAuthorMark{1}, S.~H\"{a}nsel, C.~Hartl, M.~Hoch, N.~H\"{o}rmann, J.~Hrubec, M.~Jeitler, G.~Kasieczka, W.~Kiesenhofer, M.~Krammer, D.~Liko, I.~Mikulec, M.~Pernicka, H.~Rohringer, R.~Sch\"{o}fbeck, J.~Strauss, A.~Taurok, F.~Teischinger, W.~Waltenberger, G.~Walzel, E.~Widl, C.-E.~Wulz
% ...
\vskip\cmsinstskip
% ...
1:~~Also at CERN, European Organization for Nuclear Research, Geneva, Switzerland\\
% ...
